\documentclass[twocolumn,showpacs,preprintnumbers,amsmath,superscriptaddress,amssymb]{revtex4}
\usepackage{graphicx}
\begin{document}
\title {Interaction broadening of Wannier functions and Mott transitions in atomic BEC}
\author{Jinbin Li}
\author{Yue Yu}
\affiliation{Institute of Theoretical Physics, Chinese Academy of Sciences, P.O. Box 2735, Beijing 100080, China}
\author{Artem M. Dudarev}
\affiliation{ Department of Physics, University of Texas, Austin, Texas 78712-1081, USA}
\affiliation{ Max-Planck-Institut f\"ur Physik komplexer Systeme, N\"othnitzer Str. 38, 01187 Dresden, Germany}
\author{Qian Niu}
\affiliation{ Department of Physics, University of Texas, Austin, Texas 78712-1081, USA}
\date{\today}
\begin{abstract}
Superfluid to Mott-insulator transitions in atomic BEC in optical
lattices are investigated for the case of number of atoms per site
larger than one. To account for mean field repulsion between the
atoms in each well, we construct an orthogonal set of Wannier
functions. The resulting hopping amplitude and on-site interaction
may be substantially different from those calculated with
single-atom Wannier functions. As illustrations of the approach we
consider lattices of various dimensionality and different mean
occupations. We find that in three-dimensional optical lattices
the correction to the critical lattice depth is significant to be
measured experimentally even for small number of atoms. Finally,
we discuss validity of the single band model.

\end{abstract}


\maketitle

\section{introduction}

Numerous many-body phenomena have been recently demonstrated with
Bose-Einstein condensates (BEC) in optical
lattices~\cite{orz,grei,grei1}. Number squeezing has been observed
with $^{87}$Rb atoms in a one-dimensional lattice of
pancake-shaped wells \cite{orz}, and superfluid to Mott-insulator
transitions have been witnessed with such atoms in
three-dimensional and one-dimensional optical lattices \cite{grei}
. Such transitions were predicted by theoretical studies based on
the Bose-Hubbard model \cite{BH} and by microscopic calculations
of the model parameters for BEC in optical lattices
\cite{jaks,Java}.

Very important question is whether it is possible to observe
superfluid to Mott-insulator transitions for the mean occupation number $n$
larger or even much larger than one? Phenomenological single band
Bose-Hubbard model indeed predicts such transitions.
Previous calculations of the model parameters $J$, hopping amplitude, and $U$,
on-site interaction, were based on the lowest band Wannier functions for
a single atom in an optical lattice. Repulsive interaction between the atoms
for $n>1$ may cause the wave function in each well to expand in all
directions, not only affecting the on-site interaction $U$~
\cite{oo} but also strongly enhancing tunneling $J$ between
neighboring wells. This is especially significant in lower
dimensional lattices with transverse potential bigger than the lattice
wells where large occupations can be achieved without substantial three-body
collisional loss. In order to provide theoretical guidance for
experimental observation of Mott transitions in such systems,
it is very important to obtain accurate critical parameters of
the lattice potential for lattice occupations beyond unity.

Here we show how to construct an orthogonal basis of Wannier
functions with mean-field atomic interactions taken into account.
We use it to obtain renormalized values of parameters $J$ and $U$,
from which critical depth of the potential $V_c$ is calculated for
various lattices of different dimensionality and mean occupation.
For the cubic optical lattice with $n=2$ or larger, our result is
noticeably larger than that calculated without taking into account
interaction. This increase is more pronounced for the anisotropic
cases with stronger lattice potentials in one or two directions.
For the case of one-dimensional lattice of pancake-shaped wells
\cite{orz} or two-dimensional lattice of tubes \cite{grei1}, our
results are several times larger than critical values calculated
from one-atom Wannier functions. This is in agreement with the
experimental findings that much higher lattice potentials are
needed to reach the transition point in such cases.

Kohn developed variational approach to calculate electronic Wannier
functions in crystals ~\cite{kohn}. We modify this
procedure by minimizing on-site energy self-consistently taking
into account interaction between atoms.

In the last section we address validity of the single-band
Bose-Hubbard model constructed with variational Wannier functions.
The conditions for the model to be valid need to be modified from
those for a single particle case since the interaction between the
particles alters the band structure substantially~\cite{wuniu}.
For the model to be valid two conditions have to be fulfilled:
(i) when the number of particles in a well changes by one the
variational Wannier function should not change significantly and
(ii) collective excitations of the atoms within each well should
be less energetically favorable than atom hopping between the
wells.

\section{Bose-Hubbard model and Wannier functions}

For bosonic atoms located in the lattice potential $V({\bf r})$ and described by boson field operators $\psi({\bf r})$, the Hamiltonian field operator is
\begin{eqnarray}
    H = \int {d{\bf r}\psi ^ \dagger ({\bf r})\left( { - \frac{{\hbar ^2 }}{{2m}}\nabla ^2  + V({\bf r}) } \right)} \psi ({\bf r}) \nonumber \\
    + \frac{1}{2}\frac{{4\pi a_s \hbar ^2 }}{m}\int {d{\bf r}} \psi ^ \dagger ({\bf r})\psi ^ \dagger ({\bf r})\psi ({\bf r})\psi ({\bf r}),
\end{eqnarray}
where $a_s$ is the atoms' scattering length and $m$ is the mass. To illustrate our methods we use as an example isotropic cubic lattice. We assume that the boson field operator may be expanded as
$\psi({\bf r})=\sum_i b_i W({\bf
r}-{\bf r}_i)$, where $b_i$ is the annihilation operator for an atom
in the Wannier state of site ${\bf r}_i$. Substituting this
expansion into the Hamiltonian we obtain
a problem of lattice bosons. We consider the case when the number of atoms
per cite $n_i$ fluctuates around average number $n$. This results in the standard
Bose-Hubbard Hamiltonian
\begin{eqnarray}
H=-J\sum_{\langle ij\rangle}b_i^\dag
b_j+\frac{U}{2}\sum_i n_i(n_i-1)
+\sum_{i}n_i I,\label{bhh}
\end{eqnarray}
where the effective on-site repulsion $U$, the hopping amplitude $J$ and the on-site single-atom energy $I$ are defined by
\begin{eqnarray}
    U & = & \frac{\partial^2 f}{\partial n^2} \label{UD}\\
    J & = & \int {d{\bf r}W^* ({\bf r})\left[ { - \frac{{\hbar ^2 }}{{2m}}\nabla ^2  + V({\bf r})} \right]} W({\bf r} + {\bf a}),\\
    I & = & \int d{\bf r} W^*({\bf r}
)\left[-\frac{\hbar^2}{2m}\nabla^2+V({\bf r})\right] W({\bf r}), \label{eq:i} \label{eq:j}
\end{eqnarray}
where $g=4{\pi}a_s\hbar^2/m$ and ${\bf a}$ is the lattice vector.
On-site energy $f$ is defined as
\begin{eqnarray}
f=nI+U_0 n(n-1)/2,\label{os}
\end{eqnarray}
with the bare on-site interaction $U_0$
\begin{eqnarray}
U_0=g\int d{\bf r}| W({\bf r})|^4. \label{U0D}
\end{eqnarray}
 We assumed that the
Wannier function does not change much for small fluctuations of
the number of atoms. Off-site interactions are also neglected.

In case of more than one atom per site the presence of other atoms does modify the Wannier function of an atom. Below we describe our strategy for its self-consistent calculation. We start with a
trial wave function localized in each well, $g({\bf r}-{\bf r}_i)$.
A Wannier function corresponding to the lowest Bloch band may be
constructed according to Kohn's transformation:
\begin{eqnarray}
&&W({\bf r})=\sum_i c_i g({\bf r}-{\bf r}_i),\label{kohn}\\
&&c_i=\int { d{\bf k}\over (2\pi)^3}\frac{e^{i{\bf k}\cdot{\bf
r}_i}}{\sqrt{G({\bf k})}},\nonumber
\end{eqnarray}
where the integral is over the first Brillouin zone and
\begin{eqnarray}
G({\bf k})&=&\sum_i \int d{\bf r} g({\bf r})g({\bf r}-{\bf r}_i)
\cos({\bf k}\cdot{\bf r}_i).
\end{eqnarray}
For an odd Wannier function, the cosine function should be replaced by
the sine function. One can show that such Wannier functions are
normalized and are orthogonal to each other for different wells.
We vary the trial function to minimize the on-site energy $f$~\cite{mini}.

We note that another method to calculate the Wannier functions
including interaction effects self-consistently may be used for
small interactions. Starting with non-linear time-independent
Gross-Pitaevskii equation
\begin{eqnarray}
-\frac{\hbar^2}{2m}\nabla^2\psi({\bf r})&+&\frac{4\pi\hbar^2
a_s}{m}|\psi({\bf r})|^2\psi({\bf r})\nonumber\\&+&V({\bf
r})\psi({\bf r})=\mu(\bf k)\psi(r), \label{GP}
\end{eqnarray}
one may calculate periodic Bloch states $u{\bf _k(r})$ defined as
\begin{eqnarray}
\psi_{k}({\bf r})=e^{i{\bf k\cdot r}}u{\bf_k(r})/\sqrt{N}.
\end{eqnarray}
by expanding them in Fourier series
\begin{equation}
    u_{\bf k}(x)=\sum_nA_n^{\bf k}e^{i2n\pi x/a}
\end{equation}
and solving nonlinear system of equations. Then, a set of Wannier wave functions
for the band in question is defined by
\begin{eqnarray}
W_m({\bf r}-{\bf a})&=&L^{-1/2}\sum_{BZ}\psi_{m,k}({\bf r}-{\bf
a})\nonumber\\ &=& L^{-1/2}\sum_{BZ}\psi_{m,k}({\bf r})e^{-i{\bf
k}\cdot{\bf a}}.
\end{eqnarray}
This procedure fails for large interactions because the bands develop loops and
become not single-valued~\cite{wuniu}.

\section{Superfluid to Mott-insulator transitions}

We consider three optical lattice systems which are relevant to
experiments: (i) isotropic three-dimensional optical lattice,
(ii) anisotropic three-dimensional lattices, and
(iii) the situation when the lattice potential is present only in
one or two directions and confinement in other directions is provided by
relatively weak harmonic trap.
Following standard practice, we will use the
lattice period $\pi/k$, atomic mass $m$, and recoil energy
$E_r=\hbar^2 k^2/2m$ as the basic units.

Three pairs of counter-propagating laser beams with wavelength $2 \pi / k$ propagating along three perpendicular directions create potential
\begin{eqnarray}
V({\bf r})=V_x \sin^2(kx) + V_y \sin^2 (ky)+ V_z \sin^2(kz).
\end{eqnarray}

{\it Isotropic cubic lattice} is created by the beam of equal
intensity. In this case $V_x = V_y = V_z = V_0$.

{\it Anisotropic cubic lattices} can be created by choosing intensity
of one or two beam to be much large than other. In this case
$V_y = V_z = V _ \perp \gg V_x = V_0$ or $ V_z = V _ \perp \gg V_y =
 V_x = V_0$. Below we study the case when $\hbar \omega _\perp \gg
 \mu$, where $\mu$ is the chemical potential of the atoms, thus the
 weak optical lattice is effectively one-dimensional or
 two-dimensional and transverse motion is frozen to the
 ground state of the transverse confinement.

Transverse motion can also be decoupled in the experimentally
 relevant case when the lattice potential is present only
 in {\it one or two directions} and atoms are confined in other
 directions by relatively weak harmonic trap:
  $V_T({\bf r}_\perp) = \frac{1}{2} m \omega_\perp^2 r^2_\perp$.

According to existing experiments, in our calculations through
this work, we choose the $^{87}$Rb atoms in $F=2,m=2$ state with
scattering length $a_s=5.8$ nm and the laser wavelength of 852 nm
for the three- and two-dimensional lattices and 840 nm for the
one-dimensional lattice. All numerical results are obtained using
21 lattice wells in each direction with periodic boundary
condition. Convergence has been checked using 41 wells for some of
the key results.

In each case we calculate parameters of the Bose-Hubbard model
based on the variational approach described in the previous section.
The critical condition for superfluid to Mott-insulator transition
has been found approximately as
\begin{eqnarray}
U/zJ= 2n+1+2\sqrt{n(n+1)}, \label{mfc}
\end{eqnarray}
where $z$ is the number of the nearest neighbor sites
\cite{sheshadri}. By substituting the parameters into the critical
condition, we can map out the critical potential strength as
a function of mean occupation.

In the following, we report our findings for isotropic and
anisotropic three-dimensional lattices, one-dimensional lattice of
pancake wells, and two-dimensional lattice of tubes.

\subsection{Isotropic cubic lattice }

In the case of isotropic cubic lattice we choose
variational trial function to be in the form $g({\bf
r})=g(x)g(y)g(z)$, with $g(u)=(1+\alpha u^2)e^{-u^2/\sigma^2}$,
where $\alpha$ and $\sigma$ are variational parameters.  Then the
Wannier function must also be of the product form $W({\bf
r})=w(x)w(y)w(z)$, with the one-dimensional functions $w(u)$ and
$g(u)$ related by the one-dimensional version of Kohn's
transformation.  All the three-dimensional integrals in
Eq.~(\ref{bhh})-(\ref{mfc}) can then be reduced to one-dimensional
ones, greatly simplifying the calculations.

\begin{figure}
\includegraphics[width=7.5cm]{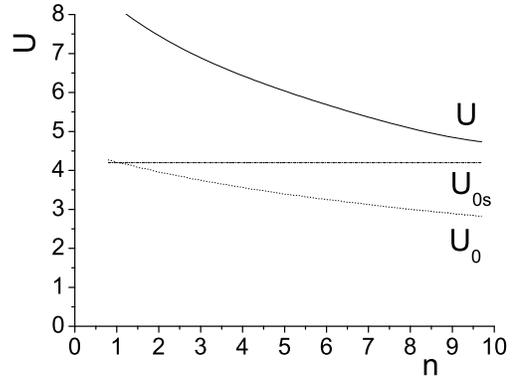}
\caption{Dependence of various interaction parameters on number of atoms for $V = 35 E_r$.
$U$ and $U_0$ are defined by (\ref{UD}) and (\ref{U0D})
respectively. The derivative in (\ref{UD}) is calculated by Chebyshev fitting to function $f$. Interaction parameter $U_{0S}$ calculated with single particle
Wannier function is defined as
$U_{0S}=g\int d{\bf r}|W_0({\bf r})|^4$, where $W_0({\bf
r})$ is a single-atom Wannier function. }\label{Ufig}
\end{figure}

\begin{figure}
\includegraphics[width=7.5cm]{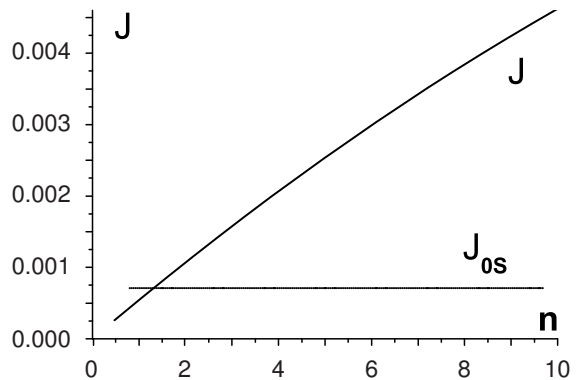}
\caption{Hopping elements calculated with single particle Wannier
function, $J_{0S}=\int {d{\bf r}W^*_0({\bf
r})\left[{-\hbar^2\nabla^2/2m+V({\bf r})}\right]}W_0({\bf r+a})$,
and with the variational proceedure described in the text, $J$.
Depth of the lattice is $V = 35 E_r$.}\label{Jfig}
\end{figure}

Our calculations proceed as following. For a given $V_0$ and $n$,
 we start with certain initial parameters $\alpha$ and $\sigma$ to
 obtain a trial Wannier function through Kohn's transformation
 and calculate the on-site energy $f$.
The procedure is repeated by varying the parameters until the
on-site energy $f$ is minimized.  The resulting variational
Wannier function will depend on both $n$ and $V_0$.  If only the
on-site single-atom energy $I$ is minimized, one obtains the
single-atom Wannier function $W_0({\bf r})$ which only depends on
$V_0$. We find that interaction broadens Wannier
functions, as a result $U_{0s}$ is always larger than $U_0$, but
we also notice that  effective interaction $U$ can be larger than
$U_0$ (see Fig.~\ref{Ufig}). So phase transition is
more complex than we expected.

\begin{figure}
\begin{center}
\includegraphics[width=7.5cm]{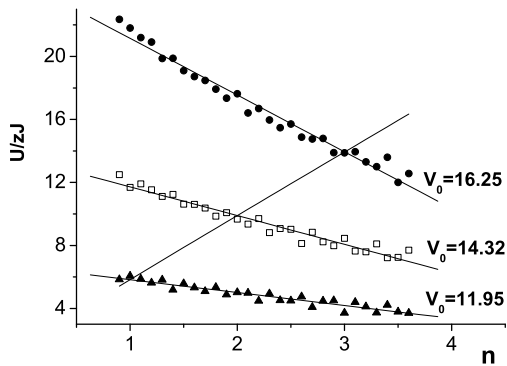}
\end{center}
\caption{\label{fig:Fig. 5} The ratio $U/zJ$ versus mean
occupation $n$ calculated from the variational Wannier functions
for isotropic cubic lattice. For each given parameter $V_0$,
intersection with the solid line yields the mean occupation number
for which the given $V_0$ is critical -- condition in
Eq.(\ref{mfc}). } \label{fig:iso}
\end{figure}

Once the Wannier function is determined, we can calculate the
Bose-Hubbard parameters $U$ and $J$. In Fig.~\ref{fig:iso}, we
depict the ratio $U/zJ$ ($z=6$) as a function of the mean
occupation $n$ for several values of the potential strength $V_0$.
The decreasing trend can be understood as following. The total
interaction energy increases with $n$, making the Wannier function
broader, hence the interaction parameter $U$ becomes smaller, $J$
proportional to overlap between neighboring Wannier functions
becomes larger, and as a result the ratio decreases. The
intersection with the line of critical condition (in
Fig.~\ref{fig:iso} the line with positive slope obtained from
Eq.~(\ref{mfc})) then yields the mean occupation for which these
potentials are critical.  For $n$=1,2,3 and 4, we find the
critical potentials to be $V_c=11.95, 14.32, 16.25$ and $18.15$
respectively. A similar calculation can be done by using the
a single-atom Wannier function.  The critical potentials become
$11.85, 13.47, 14.61$ and $15.43$ for the first four mean
occupations. For $n=1$, the two results agree with each other
within numerical uncertainty \cite{note3}, and are also consistent
with experimentally determined range for the critical potential
\cite{2}. For $n>1$, the mean field repulsion makes the critical
potential noticeably higher. Starting from $n=3$ the correction to the critical depth of the lattice has to be clearly observable experimentally and effects of interaction has to be taken into consideration.

\subsection{Anisotropic cubic lattices}

Our procedure can also be applied to the case of an anisotropic
lattice.  We model the system as a lower dimensional problem with
the reduced interaction parameter $g_d$ obtained by multiplying
$g$ by the integral of $|\psi_\perp|^4$, where $\psi_\perp$ is
the single-atom ground state wave function in a well of the
transverse potential \cite{jaks}.  In the harmonic approximation,
the wave function can be found exactly, and the reduced
interaction parameter is given by $g_1 =\frac{g
\pi}{2}\sqrt{V_{\bot}}$ for the quasi-one-dimensional lattice and
$g_2= g \sqrt{\frac{\pi}{2}}\sqrt[4]{V_{\bot}}$ for the
quasi-two-dimensional lattice.  In the calculations
discussed below, we take $V_\perp=80 E_r$.

\begin{figure}
\begin{center}
\includegraphics[width=7.5cm]{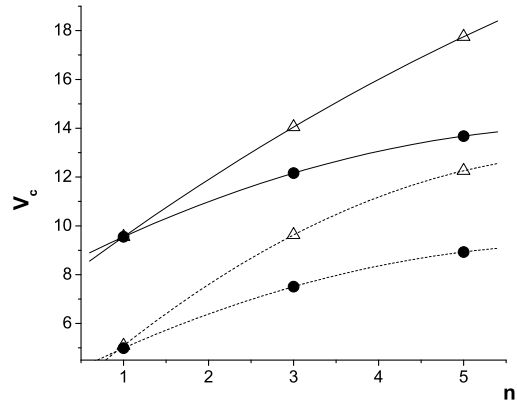}
\end{center}
\caption{\label{fig:anis} The critical lattice potential $V_{c}$
calculated from the variational and single-atom Wannier functions
for anisotropic cubic lattices. The lines are guides to eyes. The
dashed lines are for the quasi-one-dimensional and the solid lines are for
quasi-two-dimensional cases. The triangles correspond to the variational
and the circles to the single-particle calculations.}
\end{figure}

To find the Wannier functions for the lower dimensional lattices,
we use these reduced interaction parameters in our procedure,
replacing all the three-dimensional integrals in Eqs. (\ref{eq:j}) and (\ref{eq:i})
 by lower dimensional ones. The critical lattice potential
$V_c$ calculated using such variational Wannier functions is
depicted in Fig.~\ref{fig:anis} for the one- and two-dimensional models. For
comparison, we also include results calculated using the one-atom
Wannier function. The increase of critical potential due to
mean-field repulsion on the Wannier functions is somewhat bigger
in the lower dimensional cases.
\\

\begin{figure}
\begin{center}
\includegraphics[width=7.5cm]{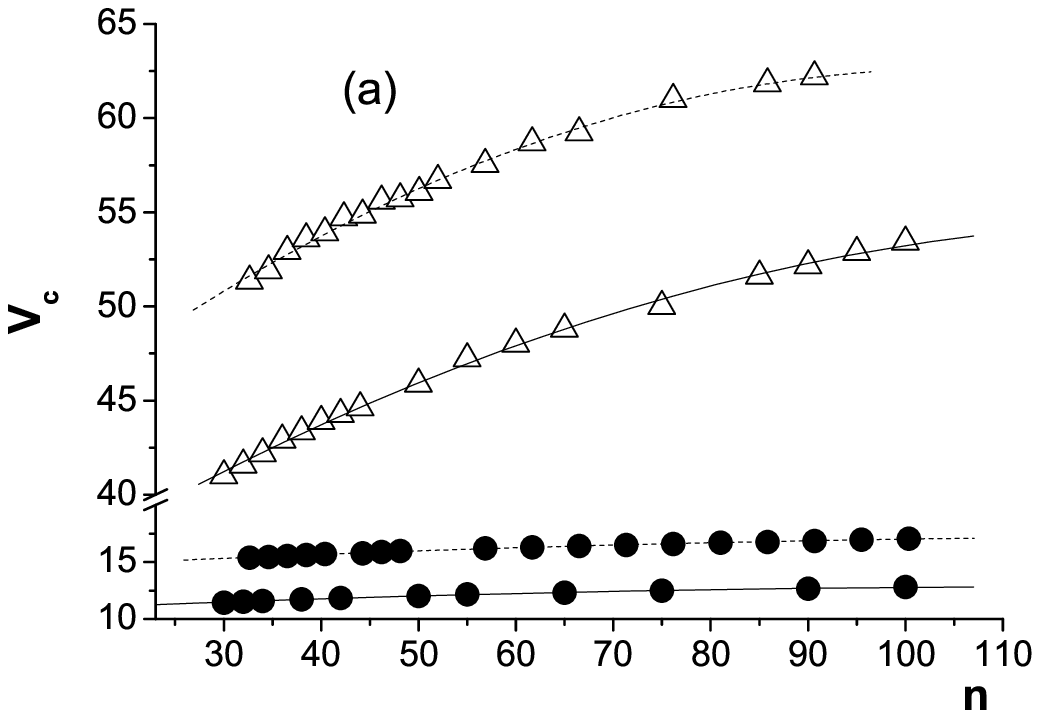}
\includegraphics[width=7.5cm]{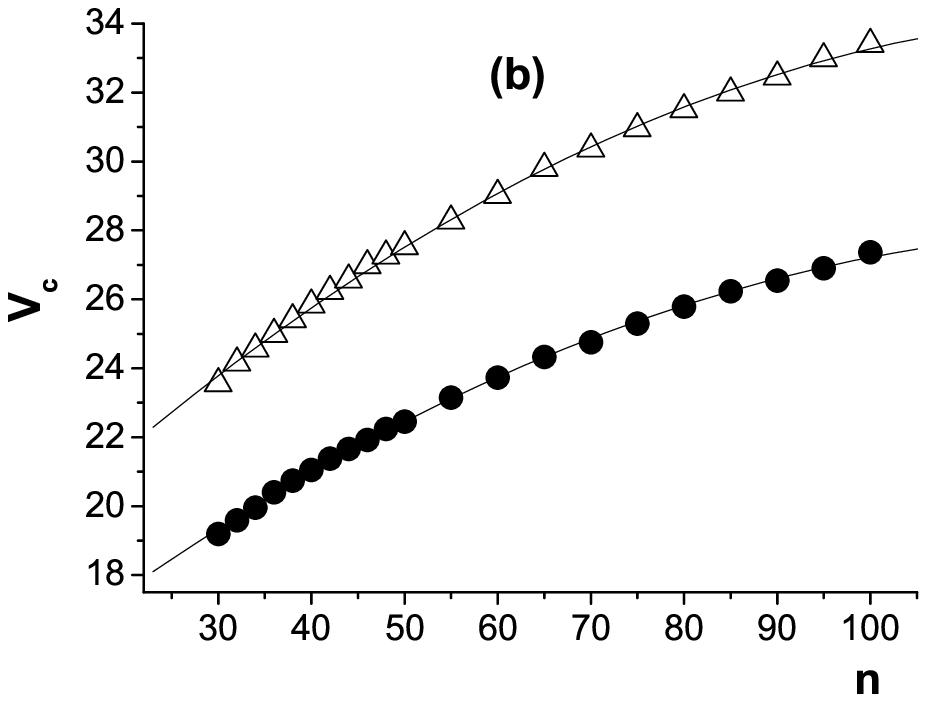}
\end{center}
\caption{\label{fig:Fig. 7} The critical lattice potential $V_c$
in dependence on mean occupation $n$ calculated from the variational (triangle)
and single-atom (circle) Wannier functions for: (a) the
one-dimensional lattice with $\omega_\perp=2\pi\times 19$ s$^{-1}$
(dashed line) and $\omega_\perp=2\pi\times 120$ s$^{-1}$ (solid line), and (b)
two-dimensional lattice with $\omega_\perp=2\pi\times 24$
s$^{-1}$.  The lines are guides to eyes.}
\end{figure}

\subsection{Lattices in one or two directions}

 BECs in
one-dimensional lattice of pancake-shaped wells and
two-dimensional lattice of tube-shaped wells have been studied in
experiments \cite{orz,3}. Because of the large transverse
dimensions of such wells, many atoms can be held in a well without
suffering too much three-atom collisional loss, opening the
possibility of studying superfluid/Mott-insulator transition for
relatively large $n$ \cite{oo,psg}. In a theoretical
investigation, Oosten et al \cite{oo} considered the interaction
effect by using a transverse wave function in the Thomas-Fermi
approximation without modifying the single-atom Wannier function
in the lattice direction(s). Here we extend their work by
considering the interaction effect on the Wannier functions as
well.

For the pancake like BEC array, the transverse wave functions are
approximated by the Thomas-Fermi wave function $\phi_{TF}({\bf
r_\perp})$ of the BEC within the pancake plane, which is defined
by
\begin{eqnarray}
|\phi_{TF}({\bf r_\perp})|^2=(ng_1)^{-1}(\mu-V_T({\bf r}_\perp)),
\end{eqnarray}
for $\mu>V_T({\bf r}_\perp)=\frac{1}{2}m\omega_\perp^2 r_\perp^2$
and vanishes otherwise. According to the experimental data, we
take $\omega_\perp=19\times 2\pi s^{-1}$.

We begin by writing the Wannier function in the form, $W({\bf r})=w({\bf r}_L)\phi({\bf r}_\perp)$, where $\phi$ is the wave function for the transverse direction(s), and $w$ is the Wannier function in the lattice direction(s), both to be determined variationally by
minimizing the on-site energy.  The part of the on-site energy
involving $\phi$ is just the $n$-particle Gross-Pitaevskii energy
in the transverse potential and with the interaction parameter $g$
modified into $g_d$ by multiplying the integral of $|w({\bf
r}_L)|^4$.  In the Thomas-Fermi approximation, this `transverse
energy' is given by $f_\perp={{2n-1}\over 3}\sqrt{n
m\omega^2_\perp g_1/\pi}$ for the one-dimensional case and
$f_\perp={{5n-2}\over 10}(9m\omega^2_\perp n^2g_2^2)^{1/3}$ for
the two-dimensional case. The total on-site energy is the sum of this
`transverse energy' and $n$ times of the single-atom energy of the
lattice Wannier function:
\begin{eqnarray}
f=f_\perp + n \int d{\bf r}_L w^*({\bf
r}_L)\left[-\frac{\hbar^2}{2m}\nabla^2+V({\bf r}_L)\right] w({\bf r}_L).
\end{eqnarray}
Lattice Wannier function, obtained by the procedure of Kohn's
transformation and minimization of the on-site energy, will be
affected by the interaction because the `transverse energy'
depends on it through the reduced interaction parameter $g_d$.
After $w({\bf r}_L)$ is determined variationally, the Bose-Hubbard
parameters $J$ and $U$ can be calculated immediately.  In Fig. 7,
we show the critical potential $V_c$ for the case of
one-dimensional lattice with transverse trap frequency
$\omega_\perp/2\pi= 19$ s$^{-1}$ and 120 s$^{-1}$.

For comparison, we also show the corresponding results obtained
using the single-atom Wannier function of the lattice and the
Thomas-Fermi transverse wave function.  It is clear that $V_{c}$
is raised dramatically due to the broadening of the Wannier
function.  In the experiment of Ref. \cite{orz}, the magnetic trap
potential is 19 s$^{-1}$. The transverse trap frequency is
enhanced to 120 s$^{-1}$ if the optical confining potential with
$V_0=50E_r$ is turned on, and the mean occupation number is $n\sim
50$. Evidence from Bragg interference pattern shows that the
critical value of the lattice potential should be somewhat larger
than $44 E_r$. This observation is contradictory to the prediction
based on the single-atom Wannier function, but is consistent with
our result based on the variational Wannier function.

In the case of two-dimensional lattice, our results for the
critical lattice potential are shown in Fig. 7(b) for
$\omega_\perp/2\pi=24$ s$^{-1}$ which is used in \cite{3}. We
predict $V_{c}\sim 33E_r$ for $n\sim 100$, while the single-atom
Wannier function yields $V_{c}\sim 27E_r$. The largest lattice
potential used in the experiment was 12 $E_r$,
so further experiment is needed to verify the theoretical predictions.\\

\section{Validity of the single-band model}

\begin{figure}[t]
\begin{center}
\includegraphics[width=7.5cm]{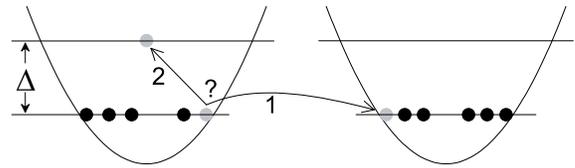}
\end{center}
\caption{\label{fig:hop} Energy associated with hopping (process 1) has to be smaller than energy to excite the many-body state in well (process 2). Many-body excitation is schematically depicted as a single atom excitation.
}
\end{figure}

\begin{figure}[t]
\begin{center}
\includegraphics[width=7.5cm]{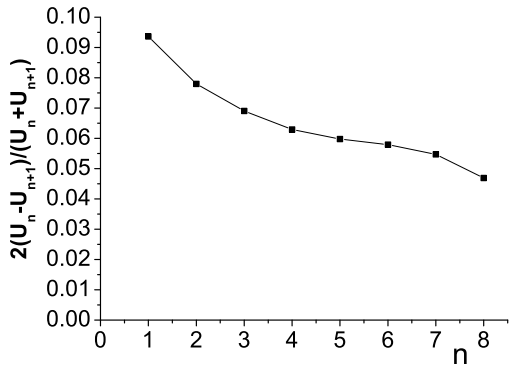}
\end{center}
\caption{Relative change in the interaction energy as number of
atoms changes by one determined by the change of the Wannier wave
 function.}
\label{fig:w_change}
\end{figure}

In this section, we discuss the conditions for the single band
Bose Hubbard model to be valid. First, we make general remarks and then give quantitative examples relevant for the case of the isotropic cubic lattice.

Assumption that the boson field operator may be expanded as
$\psi({\bf r})=\sum_i b_i W({\bf
r}-{\bf r}_i)$ requires that the Wannier functions do not change
substantially when the number of atoms in a well changes by one. A good criteria for
this condition to be fulfilled is that interaction energy calculated with
the Wannier function does not change much when number of particles changes by one
\begin{equation}
    \frac{| U_n - U_{n+1} |}{U_n + U_{n+1}} \ll 1.
    \label{eq:w_change}
\end{equation}
Note that the value of $U$ can still be quite different from the one calculated with a single particle Wannier function.

When the condition is fulfilled, the second condition is that
the excitations within the ansatz have to be
the least energetical. That is the hopping of the atoms from well has to be
more energetically favorable than excitation of atoms in each well to the
many-body excited state (see Fig.~\ref{fig:hop}). If we consider two
neighboring wells the energy of the ground state is
\begin{eqnarray}
E_0=2nI+U_0 n(n-1).
\end{eqnarray}
The energy associated with hopping is
 \begin{eqnarray}
\Delta E_1= 2nI +  U_0\frac{(n-2)(n-1)+n(n+1)}{2}-E_0=U_0.
\end{eqnarray}
It has to be much smaller than the energy of the first excited many-body state
that we denote $\Delta$
\begin{equation}
    U_0 \ll \Delta.
\end{equation}

\begin{figure}[thb]
\begin{center}
\includegraphics[width=7.5cm]{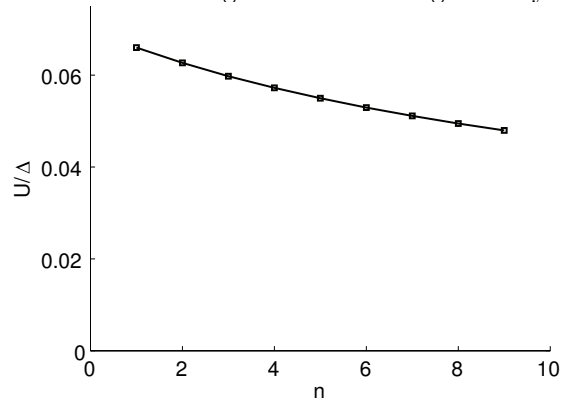}
\end{center}
\caption{\label{fig:energy} Ratio of the hopping energy to energy
required to excite atoms in each well to the lowest many-body
excited state. }
\end{figure}

We plot the criteria from Eq.~(\ref{eq:w_change}) for isotropic lattices
 on Fig.~\ref{fig:w_change}. It is much smaller than unity. To estimate
 the effect of many-body excitation within a single well, we neglect
 hopping amplitude $J$, since close to Mott-insulator transition it
 is much smaller than atom's interaction. Also for the experimentally
  relevant region of the potential depths the potential can be well
  approximated by a harmonic potential. In the harmonic potential
  the lowest many-body excited mode is associated with the center of
  mass motion -- Kohn mode~\cite{pethick}. As a result
  $\Delta \sim \hbar \omega$. Since we neglect the tunneling
  we may start directly with variational form for the Wannier
  function in a well. We take $W(x,y,z) = W(x) W(y) W(z)$, where
  $W(u) = C (1 + \beta u^2) e^{- \gamma u^2}$. Similar to previous
  section for a fixed $V_0$ and $n$ we minimize on-site energy $f$.
  From the results shown in Fig.~\ref{fig:energy} it is clear that
  the single-band model is applicable in this case: the energy
  associated with atom's hopping is much smaller than the energy
   required to excite the atoms inside of the wells.

\vspace{0.5cm}
 The authors thank Dan Heinzen, Yu-Peng Wang, Shi-Jie Yang and Li You for the
 useful
discussions. This work was supported in part by the National
Science Foundation of China, the NSF of the United States, and the
Welch Foundation of Texas.


\begin{thebibliography}{a}
\bibitem{note3} The numerical errors come from optimizing the
Wannier function, resulting in uncertainty of $\sim 0.15$ in $V_c$.
 The result for $n=1$ is slightly lower than that reported
 in \cite{2} based on single-atom Wannier function obtained
 from band structure calculations.  In principle, our calculation
 can be made more precise by using more variational parameters in the trial
 functions (See, \cite{kohn}).
\end{thebibliography}
\end{document}